\documentclass[twocolumn,letterpaper,table]{aastex6}

\usepackage{booktabs}
\usepackage{xcolor}
\usepackage{amsmath}
\usepackage{savesym}
\usepackage{txfonts}
\usepackage{mathrsfs}
\usepackage{epstopdf}
\usepackage{graphicx,color}
\usepackage{enumerate}
\usepackage{float}
\usepackage{hyperref}
\usepackage{natbib}
\newcommand{\dt}{\Delta t}
\newcommand{\ts}{t_{\rm s}}
\newcommand{\vg}{v_{\rm g}}
\newcommand{\vt}{v_{\rm t}}

\newcommand{\eqnref}[1]{Equation \ref{#1}}
\newcommand{\secref}[1]{\S\ref{#1}}
\newcommand{\figref}[1]{Figure \ref{#1}}

\begin{document}

\title{A Staggered Semi-Analytic Method for Simulating Dust Grains Subject to Gas Drag}

\author{Jeffrey Fung\altaffilmark{1,2,3}, Dhruv Muley\altaffilmark{1}}
\altaffiltext{1}{Department of Astronomy, University of California, Campbell Hall, Berkeley, CA 94720-3411}
\altaffiltext{2}{Institute for Advanced Study, 1 Einstein Drive, Princeton, NJ 08540}
\altaffiltext{3}{NASA Sagan Fellow}

\email{email: fung@ias.edu}

\begin{abstract}
Numerical simulations of dust-gas dynamics are one of the fundamental tools in astrophysical research, such as the study of star and planet formation. It is common to find tightly coupled dust and gas in astrophysical systems, which demands that any practical integration method be able to take time steps $\dt$ much longer than the stopping time $\ts$ due to drag. A number of methods have been developed to ensure stability in this stiff ($\dt\gg\ts$) regime, but there remains large room for improvement in terms of accuracy. In this paper, we describe an easy-to-implement method, the ``staggered semi-analytic method'' (SSA), and conduct numerical tests to compare it to other implicit and semi-analytic methods, including the $2^{\rm nd}$ order implicit method and the Verlet method. SSA makes use of a staggered step to better approximate the terminal velocity in the stiff regime. In applications to protoplanetary disks, this not only leads to orders-of-magnitude higher accuracy than the other methods, but also provides greater stability, making it possible to take time steps 100 times larger in some situations. SSA is also $2^{\rm nd}$ order accurate and symplectic when $\dt \ll \ts$. More generally, the robustness of SSA makes it applicable to linear dust-gas drag in virtually any context.

\end{abstract}


\section{Introduction}
\label{sec:intro}
Often in astrophysical systems, dust dynamics are strongly influenced by gas drag. Simulations of dust grains with gas drag have been used to improve our understanding of a wide range of topics, including planetesimal formation via the streaming instability \citep[e.g.,][]{Auffinger18,Rixin18,Krapp19,Umurhan19}; formation of rocky protoplanets via pebble accretion \citep[e.g.,][]{Xu17,Popovas18}; gap formation in protoplanetary disks \citep[e.g.,][]{Dong17,Dong18,Zhang19}; and dust trap in disk vortices \citep[e.g.,][]{Fu14,Zhu16,Baruteau16,Surville16,Lyra18}. Given its prevalence, it is useful to inspect further the techniques we commonly use to perform these simulations; specifically, how we integrate the motion of dust particles.

Simulating the dynamics of dust grains is known to be a challenging task, because small grains can be subjected to strong gas drag, orders of magnitude stronger than other forces such as gravity, making their equations of motion extremely stiff. While a number of methods have been used to overcome this difficulty, there remains significant room for improvement in terms of accuracy. In this work, we will introduce a new method for integrating dust motion, called the ``staggered semi-analytic method'', or SSA in short, which is a simple, easy-to-implement method that at the same time achieves high accuracy and has good conservative properties regardless of how strong the drag force is. We will compare its performance to a number of other methods, including some of the most commonly used ones.

The equation of motion for dust grains subjected to gas drag can be written as follows:
\begin{equation}
    \vec{a} = \vec{f}(t, \vec{x}, \vec{v}) + \frac{\vec{v}_{\rm g}\,(t, \vec{x}) - \vec{v}}{t_{\rm s}\,(t, \vec{x})} \, ,
    \label{eq:general}
\end{equation}
where $\vec{a}$ is the acceleration of the grain, $\vec{v}$ is its velocity, $\vec{f}$ is the specific forces acting on the dust grain except gas drag, $\vec{v}_{\rm g}$ is the gas velocity, and $t_{\rm s}$ is the ``stopping time'' that characterizes the strength of coupling between dust and gas. In this paper, we focus on linear drag, such that $t_{\rm s}$ is not itself a function of the dust velocity. This, for example, is applicable to the sub-sonic Epstein drag:
\begin{equation}
    t_{\rm s} = \frac{s}{c_{\rm T}}\frac{\rho_{\rm d}}{\rho_{\rm g}} \, ,
    \label{eq:epstein}
\end{equation}
where $s$ is the grain's size, $c_{\rm T}$ is the mean thermal speed of the gas, $\rho_{\rm d}$ is the density of the grain, and $\rho_{\rm g}$ is the density of the gas. Another example is the Stokes drag:
\begin{equation}
    t_{\rm s} = \frac{2s^2}{9\nu}\frac{\rho_{\rm d}}{\rho_{\rm g}} \, ,
    \label{eq:stokes}
\end{equation}
where $\nu$ is the kinematic viscosity.

At first glance, to obtain accurate trajectories for dust particles should require us to resolve $t_s$, i. e., use a time step $\Delta t \ll t_s$. For dust grains tightly coupled to the gas, we often find that $t_{\rm s}\ll t_{\rm dyn}$, where the dynamical time $t_{\rm dyn}$ is the characteristic timescale over which  $\vec{f}$ and $\vec{v}_{\rm g}$ change. This makes a straightforward integration over astrophysically relevant timescales prohibitively expensive; however, if we make use of the functional form of $t_s$ from \eqnref{eq:general} in constructing our numerical method, we may be able to use a $\Delta t$ that resolves only $t_{\rm dyn}$ rather than $t_s$, and thus gain orders of magnitude in efficiency.

The accuracy of a method, regardless of whether it is explicit or implicit, is related to its ``order'' --- the highest-order term in $\Delta t$ it matches with a direct Taylor expansion of the solution. When $\Delta t$ is ``small'' --- in our case, when $\tau\equiv\dt/\ts \lesssim 1$ --- higher-order methods typically yield greater accuracy. When $\tau \gtrsim 1$, however, the Taylor series diverges and the order of the method is no longer any indication of its accuracy. Explicit methods, among them the commonly used Runge-Kutta family, are unstable and completely unusable in this regime. Instead, the implicit method is often used \citep[e.g.,][]{Barriere05,Bai10,Zhu12,Laibe12b,Cuello16,Stoyanovskaya18}. Implicit methods are unconditionally stable when $\tau\gg1$, given there is no external force. Nonetheless, stability alone does not guarantee that the numerical solution is accurate. We will inspect implicit methods further later in this paper.

One attempt to get better accuracy is to exploit the fact that when $\vec{f}$, $\vec{v}_{\rm g}$, and $\ts$ are all constant, \eqnref{eq:general} has an analytic solution:
\begin{equation}
    \vec{v}~(t) = \vec{v}_{0} + \left(\vec{f}~\ts + \vec{v}_{\rm g} - \vec{v}_0\right) \left(1-e^{-t/\ts}\right) \, ,
    \label{eq:v_ana}
\end{equation}
where $\vec{v}_{0}=\vec{v}~(t=0)$. Semi-analytic methods use this expression to compute the velocity \citep[e.g.,][]{Mott00,Miniati10,Pablo15,Yang16,Rosotti16,Ishiki18}. While this certainly gives the exact solution for the velocity in one specific case, it is not immediately obvious how much of an improvement it is over implicit methods in more general cases when $\vec{f}$, $\vec{v}_{\rm g}$, and $\ts$ all change in time. In fact, there is not just one way to incorporate \eqnref{eq:v_ana} into a numerical method, and how it is done has a profound impact on the performance, as we will show.

The stiffness of \eqnref{eq:general} is one concern, but we should not neglect the case when it is not stiff. Some methods may perform well in the stiff, $\tau\gg1$, regime --- the semi-analytic method is even exact under specified conditions --- but when $\tau\ll1$, they may be easily outperformed by basic integration methods. \citet{Bai10}, for example, addresses this issue by switching between a semi-implicit and a fully implicit method. Switching between different methods is one way to adapt, but it is not ideal because the switch would inevitably produce artifacts in the numerical solution. 

We begin by describing four different methods: the first two are the $1^{\rm st}$ order implicit method and $1^{\rm st}$ order semi-analytic method, which are our baselines for comparisons; the next two are our own implementations of the $2^{\rm nd}$ order implicit method and Verlet method. These methods are comparable to some exiting methods and in some aspects similar to SSA. By comparing them against SSA, we gain insight into what truly determines the accuracy of a method when the drag is stiff. In \secref{sec:SSA}, we write down the algorithm of SSA. Following that, we perform a series of tests in \secref{sec:tests} to quantify the improvement it can provide.

\subsection{The $1^{\rm st}$ order implicit method}
\label{sec:IM1}

Implicit methods evaluate acceleration using the velocity from a future time step. For the the general equation (\eqnref{eq:general}) we want to solve, a $1^{\rm st}$ order implicit integrator, which we refer to as IM1 for convenience, can be written as follows:
\begin{align}
    v_{\rm i+1} &= v_{\rm i} + \left(f_{\rm i}~t_{\rm s,i} + v_{\rm g,i} - v_{\rm i} \right)\frac{\tau_{\rm i}}{1+\tau_{\rm i}} \, ,
    \label{eq:IM1_v} \\
    x_{\rm i+1} &= x_{\rm i} + v_{\rm i+1}~\dt \, ,
    \label{eq:IM1_x}
\end{align}
where the subscript $\rm i$ denotes values evaluated at the start of the $\rm i^{\rm th}$ time step.

When $\tau\ll1$, we get $\tau/(1+\tau)\sim \dt/\ts$. Substituting this into \eqnref{eq:IM1_v}, this implicit integrator becomes the simple $1^{\rm st}$ order symplectic Euler method \footnote{The original Euler method updates $x_{\rm i+1}$ using $v_{\rm i}$ rather than $v_{\rm i+1}$, which makes the method non-symplectic.}. When $\tau\gg1$, $v_{\rm i+1}$ has two possible limits depending on the magnitude of $v_{\rm i}$. Both limits are stable, but each with its own inaccuracies.

When $v_{\rm i}$ is small, specifically when $v_{\rm i}\ll\tau_{\rm i}v_{\rm t,i}$, where $\vt = f \ts + \vg$ is the ``terminal velocity'', $v_{\rm i+1}$ should approach $\vt$. We should keep in mind that if $f$ depends on velocity, then it is not generally possible to determine $\vt$ exactly. In physical applications, it is common to have velocity dependence in $f$, such as the centrifugal and Coriolis force. IM1 gives $v_{\rm i+1} = \vt (t_{\rm i}, x_{\rm i}, v_{\rm i})$, which has two sources of error. First, $\vt (t_{\rm i}, x_{\rm i}, v_{\rm i})$ is only a good approximation to $v_{\rm t,i}$ if the input velocity $v_{\rm i}$ is already close to $v_{\rm t,i}$. Second, this approximated $v_{\rm t,i}$ is assigned to $v_{\rm i+1}$ even though it is the terminal velocity at the start rather than the end of the step.

When $v_{\rm i}$ is large, we get $v_{\rm i+1} = v_{\rm i}/\tau_{\rm i}$. This solution guarantees stability because the particle always decelerates and the velocity does not change sign; on the other hand, it does not give the correct deceleration rate. This can be greatly improved by using a semi-analytic method.

\subsection{The $1^{\rm st}$ order semi-analytic method}
\label{sec:SA1}

The $1^{\rm st}$ order semi-analytic method, which we refer to as SA1, uses the analytic solution written in \eqnref{eq:v_ana} to update the velocity at each time step:
\begin{align}
    v_{\rm i+1} &= v_{\rm i} + \left(f_{\rm i}~t_{\rm s,i} + v_{\rm g,i} - v_{\rm i}\right) \left(1-e^{-\tau_{\rm i}}\right) \, , 
    \label{eq:SA_v} \\
    x_{\rm i+1} &= x_{\rm i} + v_{\rm i+1}~\dt \, .
    \label{eq:SA_x} 
\end{align}
Note that there is an analytic solution to the evolution of position as well, which can be written as:
\begin{equation}
    \vec{x}~(t) = \vec{x}_0 + \vec{v}_{0}~t + \left(\vec{f}~\ts + \vec{v}_{\rm g} - \vec{v}_0\right) \left(\frac{t/\ts - 1 + e^{-t/\ts}}{t/\ts}\right)t \, ,
    \label{eq:x_ana}
\end{equation}
where $\vec{x}_0 = \vec{x}~(t=0)$. One could similarly use this expression to replace \eqnref{eq:SA_x}, but in general it would not improve the numerical solution because the assumption of constant $f$, $\vg$, and $\ts$ limits the accuracy in the velocity to $1^{\rm st}$ order.

SA1 shares many similarities with IM1. When $\tau_{\rm i}\ll1$, they are both effectively the $1^{\rm st}$ order symplectic Euler method. When $\tau_{\rm i}\gg1$ and $v_{\rm i}\ll e^{\tau_{\rm i}} v_{\rm t,i}$, they both give $v_{\rm i+1} = \vt (t_{\rm i}, x_{\rm i}, v_{\rm i})$. They do differ in the limit when $\tau_{\rm i}\gg1$ and $v_{\rm i}\gg e^{\tau_{\rm i}} v_{\rm t,i}$. SA1 gives $v_{\rm i+1} = v_{\rm i} e^{-\tau_{\rm i}}$, which is almost the exact solution except for a possible spatial and temporal dependence in $\tau$.

Because SA1 is either the same or better than IM1, in later comparisons we will mainly shows results from SA1 and omit IM1.

\subsection{The $2^{\rm nd}$ order implicit method}
\label{sec:IM2}

For a $2^{\rm nd}$ order implicit integrator, we need to account for the $1^{\rm st}$ order corrections to the acceleration. For $f$, $\vg$, and $\ts$, their $1^{\rm st}$ order corrections can be accounted for by evaluating them at the middle of the time step. The drag term, $-v/\ts$, can also be generalized to $2^{\rm nd}$ order by matching the coefficients in the Taylor series of $v(t)$. Skipping the algebra, we jump to the final form of our $2^{\rm nd}$ order implicit integrator:
\begin{align}
    v_{\rm i+1} &= v_{\rm i} + \left(f_{1}~t_{\rm s,1} + v_{\rm g,1} - v_{\rm i}\right) \frac{\tau_{1}+\tau_{1}^2}{1+\frac{3}{2}\tau_{1}+\tau_{1}^2} \, , \\
    x_{\rm i+1} &= x_{\rm i} + v_{1} \dt \, ,
\end{align}
where $f_{1}$, $v_{\rm g,1}$, and $\tau_{1}$ are evaluated using the approximate midpoint values $t_{1} = t_{\rm i}+\dt/2$, $v_{1}$, and $x_{1}$:
\begin{align}
    v_{1} &= v_{\rm i} + \left(f_{\rm i}~t_{\rm s,i} + v_{\rm g,i} - v_{\rm i} \right)\frac{\tau_{\rm i}}{2+\tau_{\rm i}} \, , \\
    x_{1} &= x_{\rm i} + v_{\rm i}\frac{\dt}{2} \, .
\end{align}

We refer to the above method as IM2. When $\tau\ll1$, IM2 reduces to the midpoint method, also known as the $2^{\rm nd}$ order Runge-Kutta method. Note that among all of the methods we test in this work, IM2 is the only one that is not symplectic when the drag force is negligible. This is because the $2^{\rm nd}$ order Runge-Kutta method does not explicitly conserve the Hamiltonian of a system, unlike the symplectic Euler, Verlet, and leapfrog method which are the backbones of the other methods we tested (see their respective sections). We will later observe how this affects its performance in tests. IM2 is similar to the $2^{\rm nd}$ order fully implicit method used by \citet{Bai10}, but generalized for an arbitrary external force $f$.

When $\tau\gg1$, like IM1, IM2 also has two limits depending on the magnitude of $v_{\rm i}$. In the case when $\tau\gg1$ and $v\ll\tau v_{\rm t}$, we have $v_{1} = \vt(t_{\rm i}, x_{\rm i}, v_{\rm i})$. Assuming $v_{\rm i}$ is a good approximation to the terminal velocity at $t=t_{\rm i}$, we can write $v_{1}\sim v_{\rm t,i}$. From this, we get $v_{\rm i+1} = \vt(t_{1}, x_{1}, v_{\rm t, i})$. Note that the input parameters $x_{1}$ and $v_{\rm t, i}$ are not evaluated at the same time --- the former is evaluated at the midpoint, but the latter is evaluated at the start. This proves to be a significant source of error that we will see in later tests. Even with this discrepancy, however, IM2 should still be an improvement over IM1. Recall that IM1 gives $v_{\rm i+1}=\vt(t_{\rm i}, x_{\rm i}, v_{\rm i})$; clearly, IM2 uses better approximations for all three input parameters.

When $\tau\gg1$ and $v\gg\tau v_{\rm t}$, we get $v_{\rm i+1} = v_{\rm i}/(2\tau_{1})$. This is again an improvement over IM1, not only because it uses a better approximated $\tau$, but also because it enforces a faster deceleration that is closer to the analytic solution. It still, unfortunately, does not compare to the semi-analytic methods.

\subsection{The iterative semi-analytic Verlet method}
\label{sec:ISV}

The Verlet method is another popular method with good stability properties. It is not only an efficient method, but is also $2^{\rm nd}$ order accurate and symplectic. Moreover, because this method is explicit, it is straightforward to incorporate the semi-analytic solution given by \eqnref{eq:v_ana} to improve its accuracy. However, the Verlet method in its original form assumes the acceleration is only spatially dependent. For our problem, a direct implementation of the Verlet method would not in general lead to $2^{\rm nd}$ order accuracy. We therefore introduce here an iterative form of the Verlet method that corrects for this.

The original Verlet method reads:
\begin{align}
    v_{\rm i+1} &= v_{\rm i} + \frac{a_{\rm i} + a_{\rm i+1}}{2}~\dt \, ,
    \label{eq:Verlet_v} \\
    x_{\rm i+1} &= x_{\rm i} + v_{\rm i}~\dt + \frac{a_{\rm i}}{2}~\dt^2 \, .
    \label{eq:Verlet_x}
\end{align}
It updates the velocity using the averaged acceleration between the start and end of the time step, and updates the position using a velocity updated only to the middle of the step assuming a constant acceleration taken from the start of the step. To transform it into semi-analytic form, we can similarly use the start-to-end averaged external acceleration $f$; and instead of evaluating the acceleration due to gas drag, we apply the semi-analytic solution for the velocity, and follow the Verlet method in spirit by taking the start-to-end averages for $\vg$ and $\ts$. This way, the semi-analytic Verlet method reads:
\begin{align}
    v_{\rm i+1} &= v_{\rm i} + \left(\frac{f_{\rm i}~t_{\rm s,i} + f_{\rm i+1}~t_{\rm s,i+1}}{2} + \frac{v_{\rm g,i} + v_{\rm g,i+1}}{2} - v_{\rm i}\right) \left(1-e^{-(\tau_{\rm i}+\tau_{\rm i+1})/2}\right) \, .
    \label{eq:ISV_v} \\
    x_{\rm i+1} &= x_{\rm i} + v_{\rm i}~\dt + \left(f_{\rm i}~t_{\rm s,i} + v_{\rm g,i} - v_{\rm i}\right) \left(1-e^{-\tau_{\rm i}/2}\right)~\dt \, .
    \label{eq:ISV_x}
\end{align}
The challenge here is to evaluate $f_{\rm i+1}$ correctly. In our implementation, we first guess the value of $v_{\rm i+1}$ using \eqnref{eq:SA_v}, then use it to evaluate $f_{\rm i+1}$, and finally re-evaluate $v_{\rm i+1}$ using \eqnref{eq:ISV_v}. In our tests, we find that just one iteration as described is sufficient to bring the solution back to its expected $2^{\rm nd}$ order accuracy. We will refer to this iterative semi-analytic Verlet method as ISV.

ISV is in many ways similar to the predictor-corrector method developed by \citet{Miniati10}. Both require an estimate of $a_{\rm i+1}$ through prediction, both are $2^{\rm nd}$ order accurate, and both take advantage of the semi-analytic expression. One difference is that ISV follows the symplectic algorithm of the Verlet method for both the drag force and external force, while the predictor-corrector method treats the external force separately using a midpoint evaluation. This difference only has a minor impact on their performance.

In terms of efficiency, it appears that we have to evaluate the forces three times per step: $f_{\rm i}$ once and $f_{\rm i+1}$ twice due to the iterative step. But if the final $f_{\rm i+1}$ can be stored and reused in the next step, then there are only two evaluations, same as IM2. Moreover, since we have assumed that $\vg$ and $\ts$ are independent of velocity, they do not need to be re-evaluated during iteration, making ISV potentially more efficient than IM2.

When $\tau\ll1$, ISV has all the same properties as the original Verlet method, such as being $2^{\rm nd}$ order accurate and symplectic. When $\tau\gg1$ and $v\gg e^{\tau} v_{\rm t}$, it gives the near-exact solution like SA1, but improved because $\ts$ is now evaluated as a $2^{\rm nd}$ order accurate average over the time step. When $\tau\gg1$ and $v\ll e^{\tau} v_{\rm t}$, we get approximately $v_{\rm i+1} = \vt (t_{\rm i+1}, x_{\rm i+1}, v_{\rm t,i})$. This is similar to IM2 and shares similar inaccuracies. The underlying problem is that, since we begin with the acceleration at the start of the step, we inevitably obtain the terminal velocity at the start, leading to inconsistencies when computing external forces at any other points of the step.

ISV is already a significant improvement over IM1, IM2, and SA1. Being symplectic and $2^{\rm nd}$ order accurate when $\tau\ll1$, and semi-analytic when $\tau\gg1$, it performs well over a wide range of $\tau$. Yet, the staggered semi-analytic method presented in the following section will show even further improvement at no extra, if not less, computational cost.

\section{The Staggered Semi-Analytic Method}
\label{sec:SSA}
Here we introduce the staggered semi-analytic method (SSA), a simple method for numerically solving \eqnref{eq:general}.  The main goal is to produce a more accurate solution than all of IM1, IM2, SA1, and ISV when $\tau\gg1$, and be $2^{\rm nd}$ order accurate and symplectic like ISV when $\tau\ll1$. All the while, we keep in mind that it should be efficient and easy to implement.

To begin, we first move forward in position by half a step:
\begin{align}
    x_{1} &= x_{\rm i} + v_{\rm i}\frac{\dt}{2} 
    \label{eq:SSA_drift1}\, .
\end{align}
Then we evaluate $f$, $\vg$, and $\ts$ using this updated position, but still using the velocity from the start of the step: $f_{1} = f~(t+\frac{\dt}{2},~x_{1},~v_{\rm i})$, for example. From this we obtain the acceleration at a staggered step, where position has evolved half a step but not velocity. We then use this acceleration to estimate the midpoint velocity using the semi-analytic expression:
\begin{align}
    v_{1} &= v_{\rm i} + \left(f_{1}~t_{\rm s,1} + v_{\rm g,1} - v_{\rm i}\right) \left(1-e^{-\tau_{1}/2}\right)
    \label{eq:SSA_staggered}\, .
\end{align}

Having evaluated both $x_{1}$ and $v_{1}$, we can now use them to compute the midpoint acceleration $f_{2} = f~(t+\frac{\dt}{2},~x_{1},~v_{1})$. Finally, we finish by using this midpoint acceleration to update the velocity, and then the updated velocity to move position forward in the second half of the time step.
\begin{align}
    \label{eq:SSA_kick}
    v_{\rm i+1} &= v_{\rm i} + \left(f_{2}~t_{\rm s,1} + v_{\rm g,1} - v_{\rm i}\right) \left(1-e^{-\tau_{1}}\right) \, , \\
    \label{eq:SSA_drift2}
    x_{\rm i+1} &= x_{1} + v_{\rm i+1}~\dt/2 \, .
\end{align}
SSA evaluates $f$ twice per step, same as IM2, but $\vg$ and $\ts$ only once. This makes it as efficient as ISV, but without the need to store forces from the previous step.

In addition to being economical, SSA has a number of desirable properties. First, when $f$ is a conservative force and $\ts\rightarrow\infty$, we have $f_{1} = f_{2}$ and $1-e^{-\tau}\sim \tau$, and SSA reduces to the drift-kick-drift leapfrog method:
\begin{align}
    x_{1} &= x_{\rm i} + v_{\rm i}\frac{\dt}{2} \, , \\
    v_{\rm i+1} &= v_{\rm i} + f~\left(t_{\rm i}+\frac{\dt}{2},~x_{1}\right)~\dt \, , \\
    x_{\rm i+1} &= x_{1} + v_{\rm i+1}\frac{\dt}{2} \, .
\end{align}
In other words, SSA is a time-reversible, symplectic method in this limit. This is advantageous when integrating periodic forces over a long duration, for example. Second, when $\tau\gg1$ and $v\gg e^{\tau} v_{\rm t}$, SSA is just like ISV, giving the near-exact solution with $2^{\rm nd}$ order accuracy for $\ts$. 

Its third and most unique property is how $v_{\rm i+1}$ behaves in the other limit when $\tau\gg1$ and $v\ll e^{\tau} v_{\rm t}$. Here, we have $v_{1} = \vt(t_{\rm i}+\dt/2, x_{1}, v_{\rm i})$, or:
\begin{equation}
    v_{1} \approx v_{\rm t, i} + \frac{\partial\vt}{\partial t} \frac{\dt}{2} + \frac{\partial\vt}{\partial x} (x_{1}-x_{\rm i}) \, ,
\end{equation}
where we have again assumed $v_{\rm i}$ is a good approximation to $v_{\rm t,i}$ so that $\vt(t_{\rm i}, x_{\rm i}, v_{\rm i}) \sim v_{\rm t, i}$.
Then, $v_{\rm i+1}$ is given as $\vt(t_{\rm i}+\dt/2, x_{1}, v_{1})$, which is the midpoint terminal velocity evaluated using $v_1$, the terminal velocity at $t=t_{\rm i}$ plus corrections for the spatial and temporal dependence in $\vt$. Comparing to other methods like IM2 and ISV, we see that the staggered step provides additional corrections at no extra cost.

During review of this article, \citet{Mignone19} proposed an ``exponential midpoint'' method that shares similarities with SSA. Both methods use the drift-kick-drift leapfrog scheme as their backbones, and both modify the evaluation of $f$ in the kick step to improve the solution. While we do so using a staggered step, they integrate $f$ and average it over the time step using an exponential quadrature rule. Their test results show performance similar to SSA.

All four previously discussed methods, IM1, IM2, SA1, and ISV, are in a way similar despite their varying levels of complexity --- they all evaluate the terminal velocity at the $\rm i+1^{th}$ step using an approximated terminal velocity at the $\rm i^{th}$ step. How large is this error, and how much of it can SSA correct, are dependent on the problem and by no means obvious. Below we conduct a number of tests to quantify them.

\subsection{Implementation in spherical coordinates}
\label{sec:SSA_polar}
Often in practical applications, such as when simulating protoplanetary disks, polar coordinates are used. Here we go through the equations for SSA in spherical coordinates. It should be straightforward to reduce them to 2D polar or 3D cylindrical expressions if needed.

The main concern in spherical coordinates is the inclusion of fictitious forces. In principle, they can be included like any other velocity-dependent external forces, but one can do better by taking advantage of the conservative properties of the equations. Performance improves significantly if we evolve the angular, rather than linear, momentum equations. One finds that the Coriolis-like fictitious terms would then be absorbed, and the azimuthal angular momentum can be conserved exactly in the absence of any drag and external torque. The polar angular momentum, on the other hand, is still subject to a centrifugal term and so is not exactly conserved.

We denote $\{r,\,\theta,\,\phi\}$, as the radial, polar, and azimuthal coordinate; $\{v_{\rm r},\,j,\,l\}$ as the radial speed and the specific polar and azimuthal angular momentum; and $\{f_{\rm r},\,H,\,\Gamma\}$ as the external components of the radial force, polar torque, and azimuthal torque. \eqnref{eq:SSA_drift1} can then be written as:
\begin{align}
    r_1 &= r_{\rm i} + v_{\rm r,i}\frac{\dt}{2} \, , \\
    \theta_1 &= \theta_{\rm i} + \frac{j_{\rm i}}{r_{\rm i} r_1}\frac{\dt}{2} \, , \\
    \phi_1 &= \phi_{\rm i} + \frac{l_{\rm i}}{r_{\rm i} r_1 \sin{\theta_{\rm i}} \sin{\theta_1}}\frac{\dt}{2} \, .
\end{align}
Since the changes in $\theta$ and $\phi$ depend on the spherical radius $r$ and cylindrical radius $r\sin{\theta}$, respectively, we have to approximate these radial positions. The expressions above approximate them to $2^{\rm nd}$ order accuracy.

Next, we move to the staggered step. Same as \eqnref{eq:SSA_staggered}, $f$, $H$, and $\Gamma$, are all evaluated using $t=t_{\rm i}+\dt/2$ and the updated positions $r_1$, $\theta_1$, and $\phi_1$, but the velocities and angular momenta are still from the start of the step. These evaluations, as well as the resultant velocities and angular momenta, are given the subscript ``1''. This step goes as:
\begin{align}
    v_{\rm r,1} &= v_{\rm r,i} + \left(\left[f_{\rm r,1} + \frac{l_{\rm i}^2} {r_1^3 \sin^2{\theta_1}} + \frac{j_{\rm i}^2}{r_1^3}\right]~t_{\rm s,1} + v_{\rm r,g,1} - v_{\rm r,i}\right) \left(1-e^{-\tau_{1}/2}\right) \, , \\
    j_{1} &= j_{\rm i} + \left(\left[H_{1} + \frac{l_{\rm i}^2 \cos{\theta_1}}{r_1^2 \sin^3{\theta_1}}\right]~t_{\rm s,1} + j_{\rm g,1} - j_{\rm i}\right) \left(1-e^{-\tau_{1}/2}\right)  \, , \\
    l_{1} &= l_{\rm i} + \left(\Gamma_{1}~t_{\rm s,1} + l_{\rm g,1} - l_{\rm i}\right) \left(1-e^{-\tau_{1}/2}\right)  \, ,
\end{align}
where centrifugal forces are now taken into account.

Finally, following Equations \ref{eq:SSA_kick} and \ref{eq:SSA_drift2}, we update the solution to the $\rm i+1^{th}$ step using new values of $f$, $H$,and $\Gamma$, where they are now evaluated using $t=t_{\rm i}+\dt/2$, the updated positions $r_1$, $\theta_1$, and $\phi_1$, and the updated velocities/angular momenta $v_{\rm r,1}$, $j_1$, and $l_1$. These forces/torques are given the subscript ``2''. This update goes as:
\begin{align}
    v_{\rm r,i+1} &= v_{\rm r,i} + \left(\left[f_{\rm r,2} + \frac{l_{1}^2} {r_1^3 \sin^2{\theta_1}} + \frac{j_{1}^2}{r_1^3}\right]~t_{\rm s,1} + v_{\rm r,g,1} - v_{\rm r,i}\right) \left(1-e^{-\tau_{1}}\right) \, , \\
    j_{\rm i+1} &= j_{\rm i} + \left(\left[H_{2} + \frac{l_{\rm 1}^2 \cos{\theta_1}}{r_1^2 \sin^3{\theta_1}}\right]~t_{\rm s,1} + j_{\rm g,1} - j_{\rm i}\right) \left(1-e^{-\tau_{1}}\right)  \, , \\
    l_{\rm i+1} &= l_{\rm i} + \left(\Gamma_{2}~t_{\rm s,1} + l_{\rm g,1} - l_{\rm i}\right) \left(1-e^{-\tau_{1}}\right)  \, , \\
    r_{\rm i+1} &= r_{1} + v_{\rm r,i+1}\frac{\dt}{2} \, , \\
    \theta_{\rm i+1} &= \theta_{1} + \frac{j_{\rm i+1}}{r_{\rm i+1} r_1}\frac{\dt}{2} \, , \\
     \phi_{\rm i+1} &= \phi_{1} + \frac{l_{\rm i+1}}{r_{\rm i+1} r_1 \sin{\theta_{\rm i+1}} \sin{\theta_1}}\frac{\dt}{2} \, .
\end{align}
Note that here the centrifugal forces are evaluated using the updated (those with subscript ``1'') positions and angular momenta. This set of equations are used in \secref{sec:drift}, \secref{sec:trap}, and \secref{sec:ecc}, where simulations are done in 2D polar coordinates by setting $\theta=\pi/2$ and $j=H=0$.
\section{Tests}
\label{sec:tests}

\subsection{Deceleration test}
\label{sec:simple_drag}

\begin{figure}
    \centering
    \includegraphics[width=0.49\textwidth]{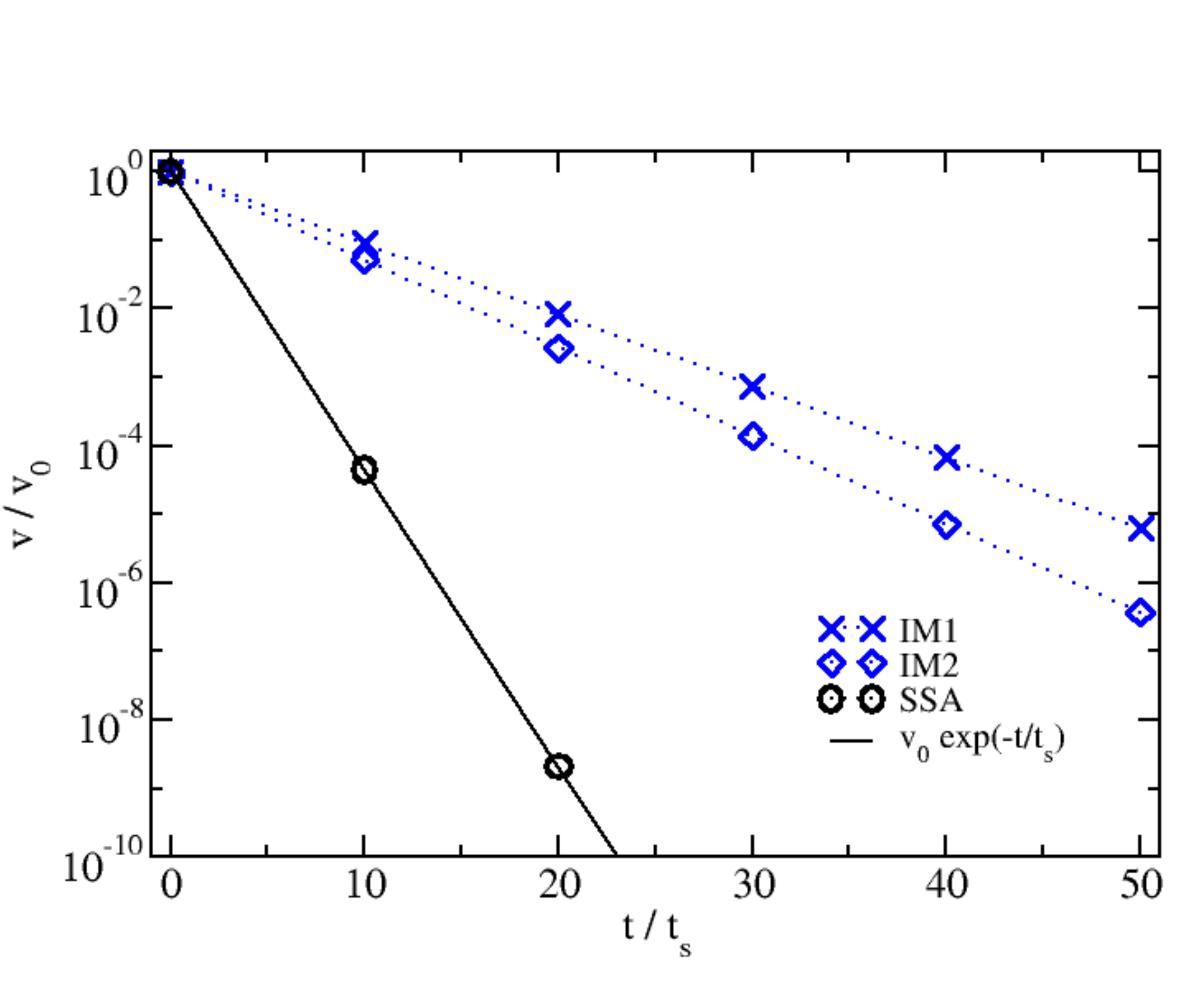}
    \caption{Particle velocity as a function of time. The black solid line is the analytic solution. The step size $\dt$ is $10~\ts$. Semi-analytic methods such as SSA (black circles) follow the analytic solution exactly, while implicit methods (blue crosses and diamonds) do not produce the correct decelerate rate.}
    \label{fig:simple_drag}
\end{figure}

To begin, we demonstrate the advantage of semi-analytic methods in a constant deceleration test. In this test, we set $f=0$, $\vg=0$, and $\ts=1$, in code units. The particle is initiated with $x=0$ and $v=1$, and we use a time step of $\dt = 10$.

\figref{fig:simple_drag} plots the velocity as a function of time computed by each of the three methods, IM1, IM2, and SSA. Not surprisingly, SSA follows the analytic solution exactly. In comparison, IM1 and IM2 both decelerates exponentially, but significantly more slowly than expected. IM2 performs only slightly better than IM1. All semi-analytic methods, such as SA1 and ISV, will perform like SSA.

This is a special case that highlights a weakness in the implicit methods. In other applications, such as in the following tests, the difference between implicit and semi-analytic methods is less extreme, but still noticeable.

\subsection{Periodic background flow}
\label{sec:periodic}

\begin{figure*}
    \centering
    \includegraphics[width=0.99\textwidth]{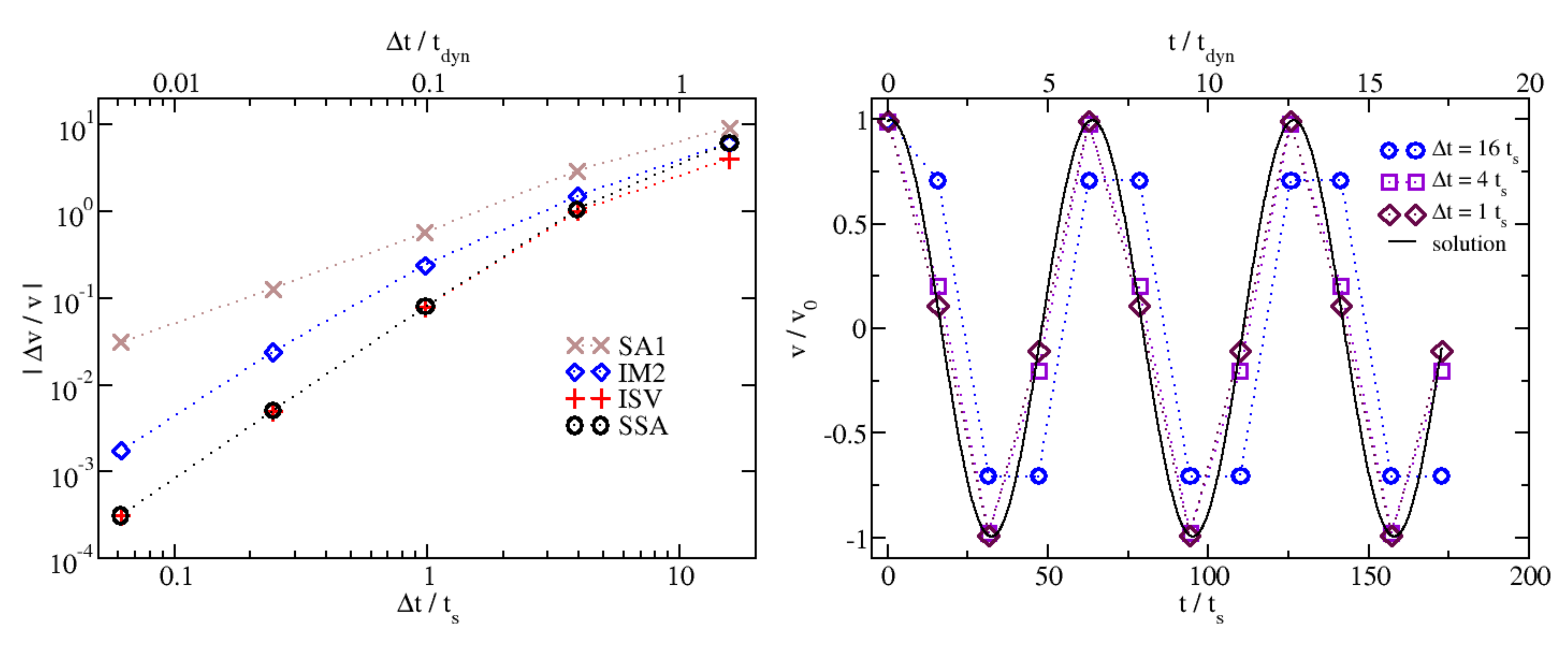}
    \caption{Results from the periodic background flow test described in \secref{sec:periodic}. The left panel plots the relative error in the velocity, $|\Delta v / v| = |(v - v_{\rm ana}) / v_{\rm ana}|$, where $v_{\rm ana}$ is the analytic solution given by \eqnref{eq:periodic}. The right panel plots velocity as a function of time computed using SSA, where we overlay \eqnref{eq:periodic} as the black solid line. The two $2^{\rm nd}$ order semi-analytic methods, ISV and SSA, perform nearly identically in this test.}
    \label{fig:periodic}
\end{figure*}

In this test, we set $f=0$, $\vg=v_0 \cos{(t/t_{\rm dyn})}$, and $\ts=1$, where $v_0 = 1$ and $t_{\rm dyn} = 10$. This problem has an analytic solution, which, in the equilibrium state, can be written as:
\begin{equation}
    v(t) = v_0\frac{\ts t_{\rm dyn} \sin{(t/t_{\rm dyn})} + t_{\rm dyn}^2 \cos{(t/t_{\rm dyn})}}{\ts^2 + t_{\rm dyn}^2} \, .
    \label{eq:periodic}
\end{equation}
We use this expression to initialize our simulations and compute the error in our numerical solutions.

We test a range of $\dt$ to measure the convergence properties of our methods. We choose $\dt = \{1/4, 1/16, 1/64, 1/256, 1/1024\}\times 2\pi t_{\rm dyn}$; which is about $\dt \sim \{16, 4, 1, 1/4, 1/16\}\times \ts$. We run the simulations for a duration of $\frac{11\pi}{2}~t_{\rm dyn}$, and then compare the velocities at that time with the analytic solution given by \eqnref{eq:periodic}. \figref{fig:periodic} illustrates our results.

In the left panel of \figref{fig:periodic}, we plot the relative error in velocity as a function of $\dt$. In the right panel, we show the numerical solution from SSA for three different time steps and compare them with \eqnref{eq:periodic}. When $\dt \ll t_{\rm dyn}$, we find the expected convergence behavior: the error in SA1 scales as $\dt^1$, while the errors in IM2, ISV, and SSA scale as $\dt^2$. When $\dt \gtrsim t_{\rm dyn}$, the time step becomes too coarse to adequately sample the change in $\vg$, and the error in all four methods converge to each other. There is no clear change in behavior around $\dt \sim \ts$. All four methods are therefore capable of taking large time steps for this type of stiff equation. 

The error in SSA is about the same as ISV, and 5 times smaller than IM2. Because the $f$ does not depend on velocity in this test, the staggered step in SSA is no different from a simple midpoint evaluation of the acceleration. Therefore, the difference between SSA and IM2 mainly demonstrates the benefits of utilizing the semi-analytic solution (\eqnref{eq:v_ana}). The fact that SSA and ISV perform nearly identically confirms this.

\subsection{Dust drift in a circumstellar disk}
\label{sec:drift}

\begin{figure*}
    \centering
    \includegraphics[width=0.99\textwidth]{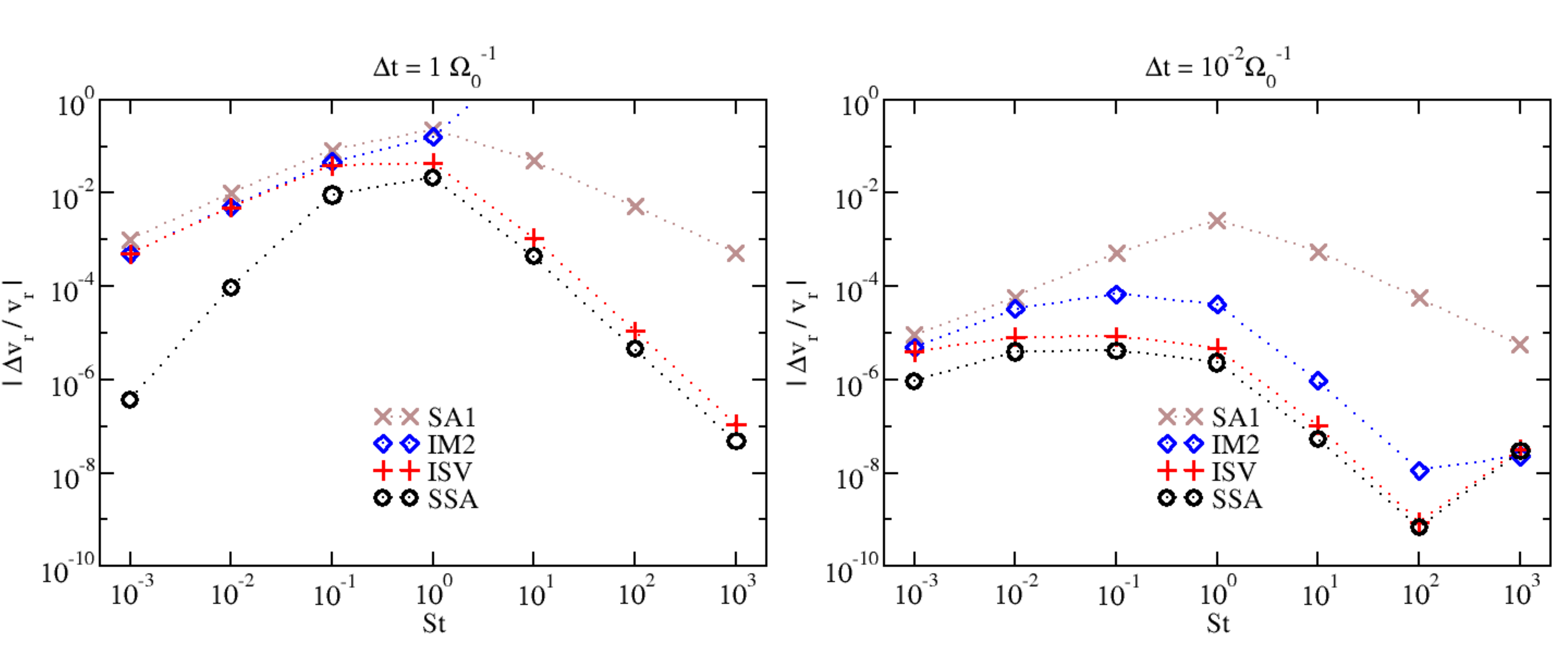}
    \caption{Results from the dust drift test described in \secref{sec:drift}. We plot the relative error in the drift velocity for four different methods when $\dt = 1~\Omega_0^{-1}$ on the left, and $\dt = 10^{-2}~\Omega_0^{-1}$ on the right. Error is measured against the approximated solution given by \eqnref{eq:drift_vr}. The staggered step in SSA eliminates the inconsistency between the evaluation of the centrifugal force and the terminal velocity, which leads to higher accuracy than the other three methods across the entire range of $\mathit{St}$, but especially when $\mathit{St}<\dt \, \Omega_0$. }
    \label{fig:drift}
\end{figure*}

\begin{figure}
    \centering
    \includegraphics[width=0.49\textwidth]{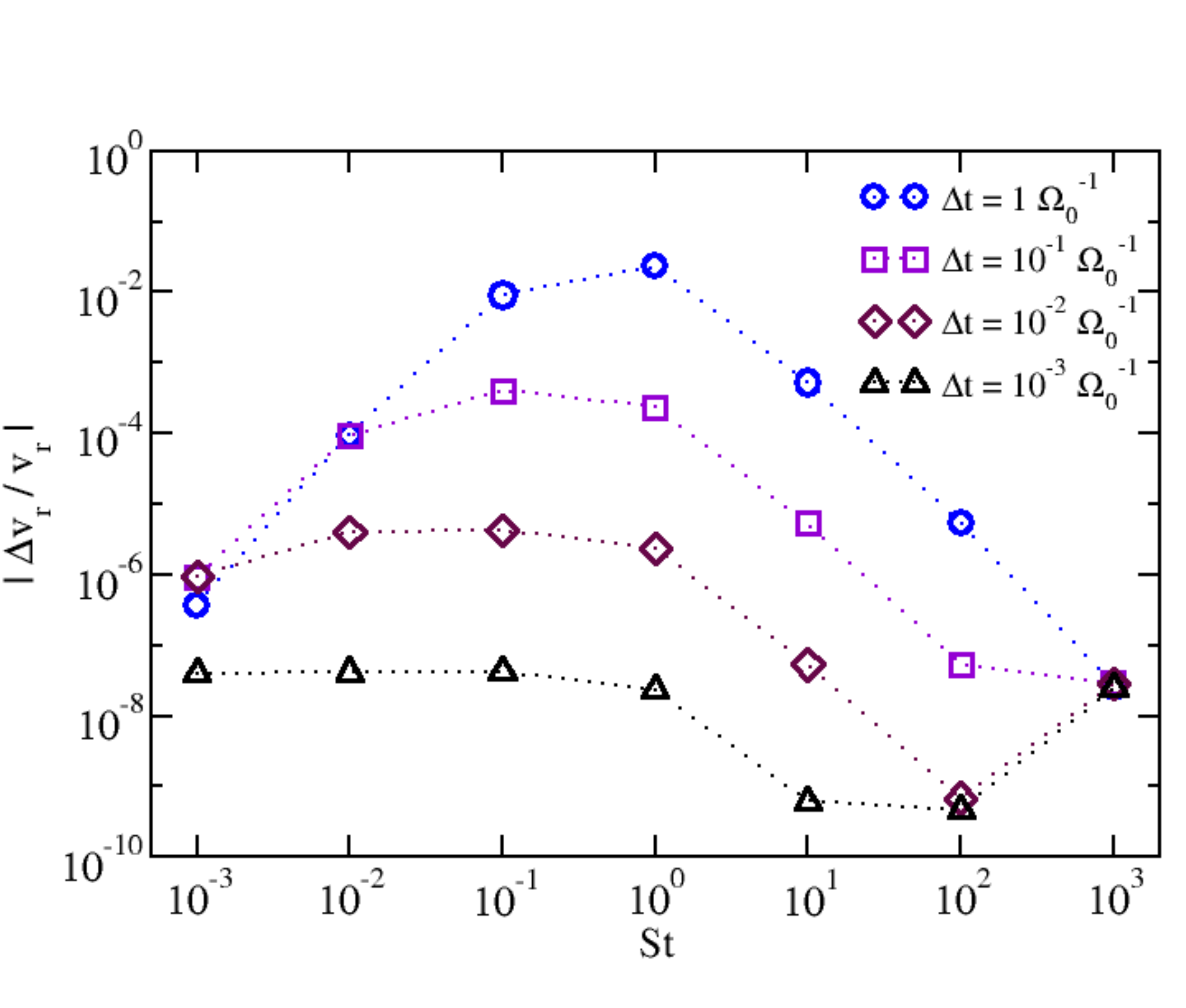}
    \caption{Similar to \figref{fig:drift}, but only for SSA to demonstrate its convergence properties. Here the precision in \eqnref{eq:drift_vr} is not sufficient for measuring errors as small as $10^{-10}$, so they are instead measured against the numerical solution given by SSA using $\dt = 10^{-5}\Omega_0^{-1}$. Error in SSA mostly scales with $\dt^{2}$, except when $\mathit{St}\gg1$, where it becomes dominated by roundoff error with decreasing $\dt$, or when $\mathit{St}\ll1$, where it plateaus to a constant with increasing $\dt$.}
    \label{fig:drift_conv}
\end{figure}

In this test, we place particles inside a gaseous disk orbiting a central mass. The simulations are performed in 2D polar coordinates, denoting the radial and azimuthal coordinates as $\{r,\phi\}$. \secref{sec:SSA_polar} describes in detail how we implement SSA in this coordinate system; similar transformations are done for the other methods. The disk is assumed to be in equilibrium, with gravity balanced by pressure and centrifugal force. The orbital speed of the disk can then be written as:
\begin{equation}
    v_{\phi,\rm g} = v_{\rm K} \sqrt{1 + H^2 \left( \frac{{\rm d}\ln{c_{\rm s}^2}}{{\rm d}\ln{r}} + \frac{{\rm d}\ln{\Sigma}}{{\rm d}\ln{r}}\right)} \, ,
    \label{eq:v_gas}
\end{equation}
where $v_{\rm K}$ is the Keplerian speed, $c_{\rm s}$ is the sound speed in the disk, $\Sigma$ is the disk surface density, and $H = c_{\rm s}/v_{\rm K}$ is the disk aspect ratio. For this test, we choose $H = 0.05$, $\frac{{\rm d}\ln{c_{\rm s}^2}}{{\rm d}\ln{r}} = -1$, and $\frac{{\rm d}\ln{\Sigma}}{{\rm d}\ln{r}} = 0$. The radial speed of the gas, $v_{\rm r, g}$ is assumed to be zero.

The forces acting on the particles other than gas drag are gravity and centrifugal force:
\begin{equation}
    f = -\frac{GM}{r^2} + \frac{l^2}{r^3} \, ,
\end{equation}
where $G$ is the gravitational constant, $M$ is the central body's mass, and $l = r v_{\phi}$ is the specific angular momentum of the particle. The stopping time $\ts$ is defined using the Stokes number $\mathit{St}$, such that:
\begin{equation}
    \ts = \mathit{St}~\Omega_{\rm K}^{-1} \, ,
    \label{eq:drift_ts}
\end{equation}
where $\Omega_{\rm K} = \sqrt{GM / r^3}$ is the Keplerian orbital frequency. We also denote $r_0$ as the starting position of the particle, and $\Omega_0$ as $\Omega_{\rm K}$ evaluated at $r_0$.

Because the particles do not feel pressure force, they tend to orbit at Keplerian speed, whereas the gas orbits at a sub-Keplerian speed due to the presence of some pressure support. Gas drag therefore exerts a negative torque on the particle that leads to an inward radial drift. We can define $l = l_{\rm K} (1-L)$ as the particle's angular momentum, where $l_{\rm K} = \sqrt{GMr}$, and $L$ is a dimensionless variable. Then the radial drift speed is approximately $f\ts$, or:
\begin{equation}
    v_{\rm r} \approx -2L ~\left(1+\frac{L}{2}\right) ~\mathit{St} ~v_{\rm K} \, ,
    \label{eq:drift_vr}
\end{equation}
and for our setup, it can be shown that in equilibrium state:
\begin{equation}
    L \approx  \frac{1 - \sqrt{1-H^2}}{1 + \mathit{St}^2} \left(1 + \frac{3 \mathit{St}^2}{2}\frac{\left(1 - \sqrt{1-H^2}\right)}{\left(1 + \mathit{St}^2\right)^2}\right)\, .
    \label{eq:drift_L}
\end{equation}
To a good approximation, $L$ can be more simply written as $(H^2 / 2)/(1 +\mathit{St}^2)$, and  $v_{\rm r}$ as $-v_{\rm K} H^2 \mathit{St} / (1 +\mathit{St}^2)$, but for better precision when measuring error, we will use the more accurate form of Equations \ref{eq:drift_vr} and \ref{eq:drift_L}.

Initializing particles at $r_0$ with speeds corresponding to Equations \ref{eq:drift_vr} and \ref{eq:drift_L}, we simulate them over a duration of ${\rm max}(1, \mathit{St}) 10~\Omega_0^{-1}$, which ensures the simulations last at least 10 stopping time. We sample a range of different $\mathit{St}$ and $\dt$, and show our results in \figref{fig:drift}.

\figref{fig:drift} compares the relative error in the drift velocity as a function of $\mathit{St}$ for the different methods. For this set of simulations, we choose $\dt = 1\,\Omega_0^{-1}$ (left panel) and $10^{-2}\,\Omega_0^{-1}$ (right panel). The error in SA1 peaks at $\mathit{St}=1$, but for ISV, and SSA, the error tends to be highest when $\mathit{St}\sim\dt\,\Omega_0$. IM2 becomes unstable when the particle is decoupled from the gas ($\mathit{St}>1)$ and $\dt$ is comparable to the dynamical time $\Omega_0^{-1}$. This has to do with the fact that it lacks the symplectic property of other three. We will explore this further in \secref{sec:ecc}.

When $\dt/\ts<1$, all methods behave similar to the previous tests (except when IM2 becomes unstable). SA1 shows $1^{\rm st}$ order convergence, while IM2, ISV, and SSA are $2^{\rm nd}$ order, as expected. SSA slightly out-performs ISV by about a factor of 2, showing that when there is a velocity-dependence in the force , i.e. the centrifugal force, SSA has an advantage over ISV.

In the $\dt/\ts>1$ regime, SA1, IM2 and ISV all give close answers. IM2 and ISV are nearly identical and both are about a factor of 2 better than SA1. Their error reduces linearly with decreasing $\mathit{St}$. SSA is the only method that shows a different scaling, with error roughly scaling with $\mathit{St}^2$ instead, which leads to a three orders of magnitude smaller error than IM2 and ISV when $\mathit{St}=10^{-3}$ and $\dt=1\,\Omega_{0}$, and about one order of magnitude when $\dt=10^{-2}\,\Omega_{0}$.

To understand how SSA achieves this high accuracy in the $\dt/\ts>1$ regime, we can go through the steps taken in the SSA method. At the staggered step, the angular momentum equation sets the particle's angular momentum to its terminal value at the midpoint position. Following that, the midpoint step uses this angular momentum to evaluate the centrifugal force at the midpoint, thus producing the correct radial velocity at the midpoint. It is particularly ideal in setup, because the terminal value of the angular momentum is only a function of space defined by the angular momentum of the background gas, meaning that the angular momentum given by the staggered step is the exact solution. SSA therefore outstrips all others in terms of accuracy, and, as we will see in the following, stability as well.

We examine how the error in SSA changes with $\dt$ in \figref{fig:drift_conv}. As expected, its error scales with $\dt^2$ in most cases, with two exceptions. When $\mathit{St}>10^2$, the relative error becomes dominated by roundoff error at the shortest timesteps. This is because we have reached machine precision in terms of the absolute error in the particle's angular momentum. To reach a relative error less than $10^{-10}$ in $v_{\rm r}$, angular momentum has to be accurate to a level of $10^{-10}~L\sim 10^{-8} (H^2/\mathit{St}^2) \sim 10^{-16}$, but a double precision float variable only keeps track of 15 significant digits. 

On the other end, we find when $\dt/\ts>1$, the error does not increase beyond a particular value, even when $\dt/\ts$ is as large as $10^3$. Specifically, the relative error tops out at $\sim10^{-6}$ when $\mathit{St}=10^{-3}$, and $\sim10^{-4}$ when $\mathit{St}=10^{-2}$. Evidently, the combined effects of the semi-analytic solution and an accurate evaluation of the centrifugal force using the staggered step allows SSA to produce a near-exact solution regardless of the size of the time step. The property implies the stability of SSA is much beyond the other methods, and should allow us to take very large time steps in physical applications. Our next test will verify this.

\subsection{Dust trap in a circumstellar disk}
\label{sec:trap}

\begin{figure}
    \centering
    \includegraphics[width=0.49\textwidth]{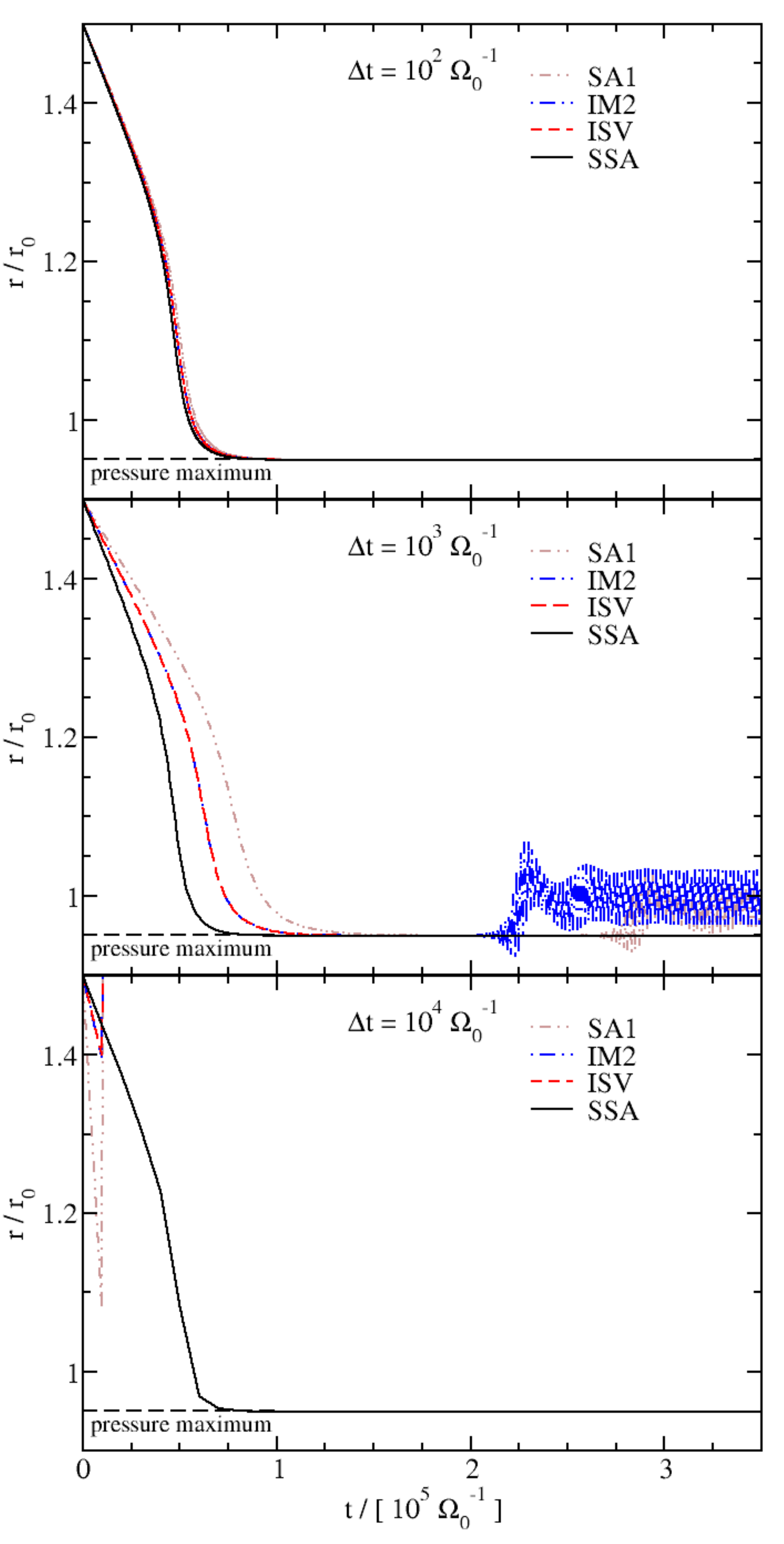}
    \caption{Particle positions as functions of time from the dust trap test described in \secref{sec:trap}. Particles are expected to drift inward and become trapped at the pressure maximum located at $r\sim0.95 r_0$, labeled by the black dashed lines. The three panels show results using different time steps, ranging from $\dt = 10^2 \Omega_0^{-1}$ in the top panel to $10^4 \Omega_0^{-1}$ in the bottom panel. In all three panels, SSA consistently computes the same drift speed and captures the stalling at the pressure maximum. The other three methods can only reproduce its results using the shortest time step, $10^2 \Omega_0^{-1}$. SA1 and IM2 starts to exhibit unstable behavior in the middle panel when $\dt = 10^3 \Omega_0^{-1}$; and all methods except SSA break down within two steps in the bottom panel when $\dt = 10^4 \Omega_0^{-1}$.}
    \label{fig:trap}
\end{figure}

In this test, we use the same setup as the previous test, but include a dust trap in the background gas disk by introducing a bump in its surface density profile:
\begin{equation}
    \Sigma = \Sigma_0 + \Sigma_1 \exp{\left[-\frac{(r-r_0)^2}{2w^2}\right]} \, ,
\end{equation}
where we choose $\Sigma_0 = 1$, $\Sigma_1 = 0.3$, $r_0 = 1$, and $w=0.1$. Plugging this profile into \eqnref{eq:v_gas}, we get $v_{\phi,\rm g} = v_{\rm K}$ at $r\sim0.95 r_0$. The gas orbital speed is sub-Keplerian outside of this radius, and super-Keplerian inside, making it a trap for our particles. Physically, it corresponds to the location of the disk's local pressure maximum.

Similar to the previous test, we initiate particles according to Equations \ref{eq:drift_vr} and \ref{eq:drift_L}, but with starting positions at $1.5 r_0$. We fix their $\mathit{St}$ to be $10^{-3}$, and compute $\ts$ according to \eqnref{eq:drift_ts}. We choose three different time steps: $\dt = \{10^2, 10^3, 10^4\}~\Omega_0^{-1}$. Since the drift speed is about $v_{\rm r}/v_{\rm K}\sim H^2 \mathit{St} \sim 10^{-6}$, the change in radial position per step, $\Delta r$, is $\Delta r/r \sim \{10^{-4}, 10^{-3}, 10^{-2}\}$, respectively. Considering how slow the drift is, these large time steps are reasonable if one wishes to simulate this evolution efficiently.

These particles should first drift toward the pressure maximum, and then stall as they approach it. In the top panel of \figref{fig:trap}, all 4 methods, SA1, IM2, ISV, and SSA, correctly simulate this motion when $\dt = 10^2~\Omega_0^{-1}$. When we increase the step size by a factor of 10 to $10^3~\Omega_0^{-1}$, SA1 and IM2 both show some unstable behavior, shown in the middle panel of \figref{fig:trap}. As the particles decelerate toward the pressure maximum, they gain some excess angular momentum from numerical inaccuracy, and go into eccentric orbits that the large time step cannot correctly track. ISV and SSA, on the other hand, remain stable, although the drift speed given by ISV errs on the slow side by $\sim 20\%$. When we increase the step size by yet another factor of 10, we see in the bottom panel of \figref{fig:trap} that SA1, and IM2, and ISV can no longer track dust drift correctly, producing large errors within the first two steps. Impressively, SSA remains robust. This reflects the same behavior we observed in the previous test, where we find that the error in SSA has an upper limit when $\tau\gg1$.

The ability to take time steps as large as $10^4~\Omega_0^{-1}$ makes SSA stand out among all the methods we tested. This opens up the possibility of testing analytic models of dust evolution in protoplanetary disks \citep[e.g.][]{Pinilla16,Cridland17,Pinilla17,Sierra19,Garate19} with simulations of dust dynamics. These models typically require the addition of diffusion physics (due to disk turbulence), which we have not discussed in this work. Diffusion can be included more easily in an Eulerian, grid-based simulation. Our description of SSA has so far been a Lagrangian, particle-based approach, but it is possible to apply it to a grid setting as well. \secref{sec:fluid} will discuss this topic in more detail.

\subsection{Eccentricity damping}
\label{sec:ecc}

\begin{figure}
    \centering
    \includegraphics[width=0.49\textwidth]{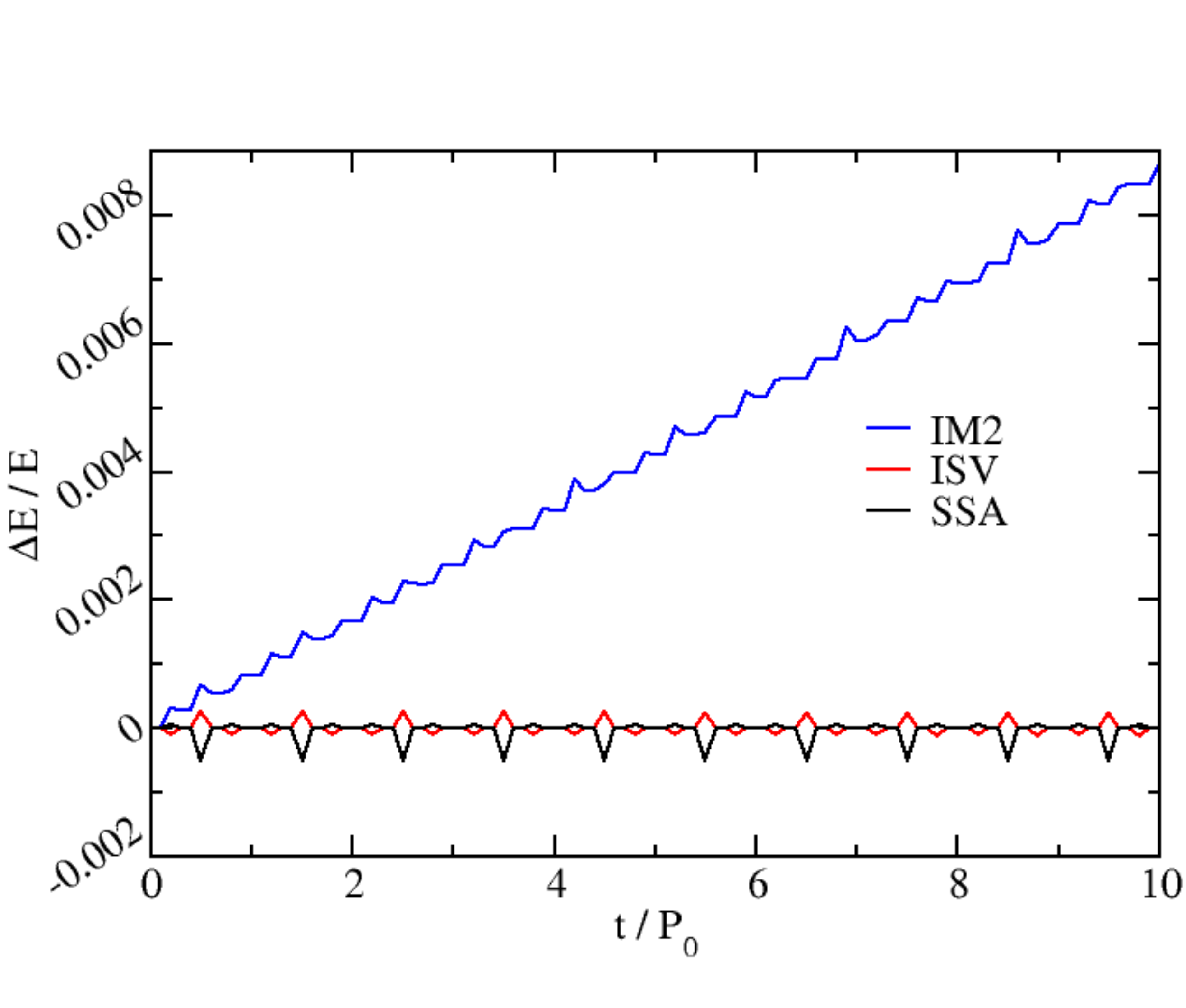}
    \caption{The error in the energy of a particle on an $e=0.5$ eccentric orbit. $\mathit{St}$ is set to be $10^{15}$ so that drag is negligible. ISV and SSA can conserve energy in the long-term because they are symplectic methods in this limit. IM2, on the other hand, accumulates error and is not suitable for this type of integration.}
    \label{fig:ecc_E}
\end{figure}

\begin{figure*}
    \centering
    \includegraphics[width=0.99\textwidth]{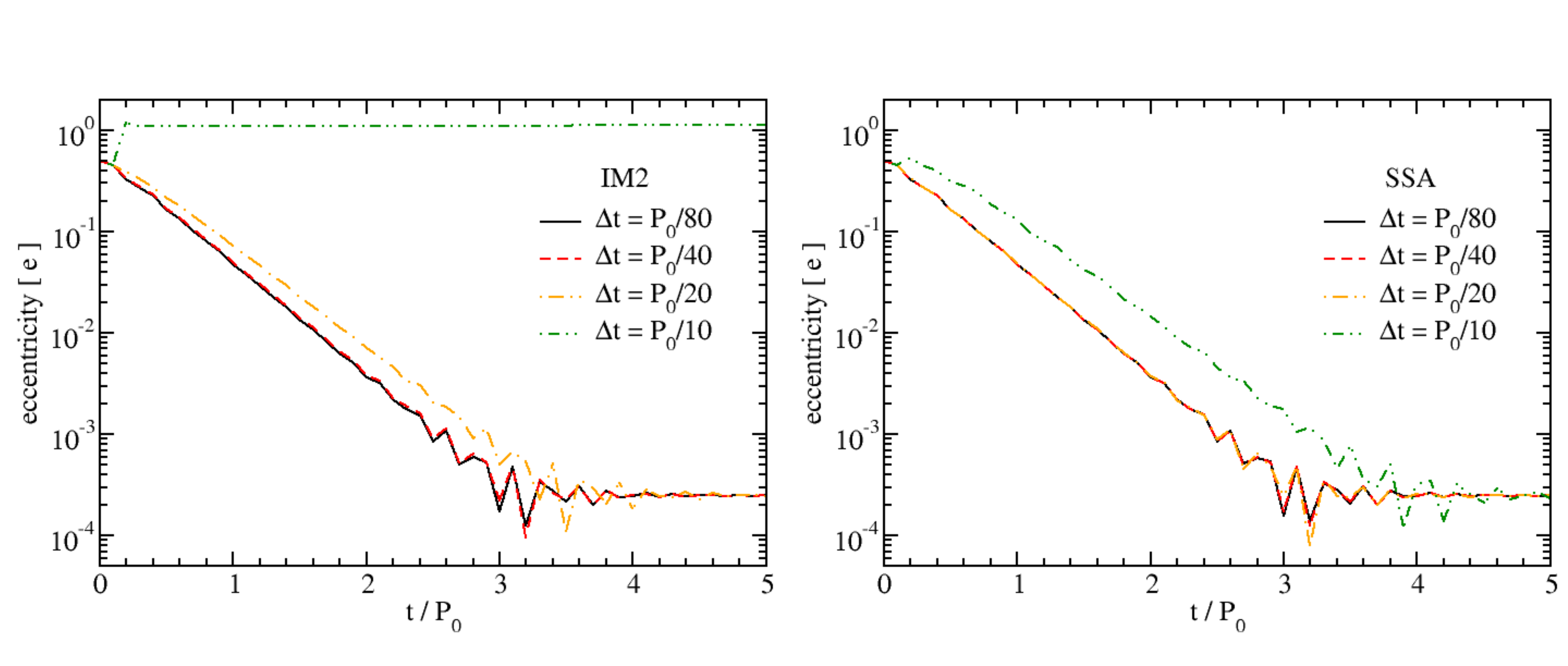}
    \caption{Results from the eccentricity damping test described in \secref{sec:ecc}. We plot the particle's eccentricity as a function of time, and show the results for IM2 on the left, and SSA on the right. $\mathit{St}$ is set to be $10$, so the stopping time is approximately 2 orbital periods, $P_0$. SSA converges faster and is more stable than IM2, despite both are $2^{\rm nd}$ order methods and the drag is not stiff. This is because SSA is symplectic, thus can track eccentric orbits more accurately, as shown in \figref{fig:ecc_E}.}
    \label{fig:ecc_s1}
\end{figure*}

We once again take a similar setup as \secref{sec:drift}, but this time we initiate particles with eccentricities of $0.5$. In this case, symplectic methods, which conserve the Hamiltonian of particles, have a significant advantage in tracking the orbits of the particles over many orbital periods.

In \figref{fig:ecc_E}, we plot $\Delta E$, the change in the particle's energy from their initial values as a function of time. The particles are given $\mathit{St}=10^{15}$ so that drag is negligible, and the time step $\dt$ is $P_0/160$, where $P_0 = 2\pi \Omega_0^{-1}$ is the orbital period of the particle. As mentioned in Sections \ref{sec:IM2}, \ref{sec:ISV}, and \ref{sec:SSA}, in this limit IM2 is equivalent to the $2^{\rm nd}$ order Runge-Kutta method, ISV is equivalent to the Verlet method, and SSA is the drift-kick-drift leapfrog method. Since the latter two are symplectic, it shows in \figref{fig:ecc_E} that, while the error in $E$ is comparable between the three methods within one orbit, only the symplectic methods are stable in the long-term.

When drag is included, dissipation due to drag reduces the impact of failing to conserve energy, but does not completely remove it. In \figref{fig:ecc_s1}, we give the particles $\mathit{St}=10$ so that their eccentricities are damped over time. We find that IM2 is still unstable when the time step is as short as $P_0/10$, while SSA (also ISV which we omit because it performs nearly identically to SSA) can track eccentricity damping to a high accuracy for the full range of $\dt$ tested.

\section{Conclusion}
\label{sec:conclude}

To find the most accurate method for integrating the motion of dust particles subject to a linear drag (\eqnref{eq:general}), we have analyzed and measured the performance of five different methods, the $1^{\rm st}$ and $2^{\rm nd}$ order implicit methods (IM1 and IM2), the $1^{\rm st}$ order semi-analytic method (SA1), and two new methods --- the iterative semi-analytic Verlet method (ISV) and the staggered semi-analytic method (SSA). 

We find that all methods except SSA suffer a similar deficiency --- when the equation is stiff ($\tau=\dt/\ts\gg1$), there is a discrepancy in the temporal and spatial locations between where external forces and terminal velocities are evaluated.
By taking advantage of a staggered step in its algorithm, SSA can partially correct for this discrepancy, which leads to significantly better performance both in terms of accuracy and stability, sometimes by orders of magnitude. The improvement is particularly pronounced when simulating small dust grains embedded in protoplanetary disks (\secref{sec:drift} and \secref{sec:trap}). SSA proves to be highly robust, never failing any of our tests, regardless of how stiff the equation is. Moreover, SSA is efficient and easy to implement. We therefore recommend it as one of the best methods for integrating dust dynamics.

In this work, we have focused on linear drag, where $\ts$ is independent of velocity. It is possible to extend the algorithm of SSA to other drag regimes if there is some semi-analytic expressions one can use in place of \eqnref{eq:v_ana}. For example, when the speed of the particle is supersonic relative to the gas, the acceleration due to drag can be written as:
\begin{equation}
    \vec{a} = (\vec{\vg} - \vec{v})\frac{|\vec{\vg} - \vec{v}|}{c~\ts} \, ,
    \label{eq:super}
\end{equation}
where $c$ is a some constant with units of speed. The analytic solution for velocity in one-dimension when $\vg$ and $\ts$ are constant is:
\begin{equation}
    v\,(t) = v_0 + (\vg - v_0)\left[\frac{|\vg - v_0|}{c} \frac{t}{\ts}\right]/\left[1 + \frac{|\vg - v_0|}{c} \frac{t}{\ts}\right] \, ,
    \label{eq:super_ana}
\end{equation}
where $v_0 = v\,(t=0)$. When one includes a non-zero external force $f$, the solution becomes significantly more complicated. In the Appendix, we explore how it can be treated, but the details of its implementation, such as how to connect different drag regimes, are beyond the scope of this paper. 

In practice, it is rare that one would simultaneously be in the supersonic regime and have $\tau\gg1$, because $\tau\gg1$ usually implies dust grains are tightly coupled to the gas. If $\tau\ll1$, then the use of a semi-analytic expression is not essential. In fact, one can still employ SSA in the exact same form we described by replacing $\ts$ with $\ts\,c/|\vg-v|$. It is worth noting that \eqnref{eq:super_ana} resembles \eqnref{eq:IM1_v}. When $f=0$ and $\tau = \dt\,|\vg-v|/(c\,\ts)$, IM1 is effectively a semi-analytic method for solving \eqnref{eq:super}.

\subsection{Dust-gas hybrid implementation}
\label{sec:fluid}

SSA can be generalized to a dust-gas hybrid implementation. Here we chart the pathways to designing such a code. One way is to use SSA directly as we described to simulate Lagrangian dust particles, and include extra algorithms to follow the interaction between the particles and the grid-based hydrodynamics. For this approach, there are three main issues to consider: 1) how to interpolate the hydro grid to obtain $\ts$ and $\vg$; 2) how to distribute the transfer of momentum from dust to gas; and 3) how to obtain $\ts$ and $v_{\rm g}$ in the middle of a time step to be used in Equations \ref{eq:SSA_staggered} and \ref{eq:SSA_kick}. On the last point, it is dependent on the hydro method. Some methods already contain the midpoint information, such as if it uses a Runge-Kutta algorithm; otherwise, extra steps would need to be added to use SSA.

On the first two points, there exists a number of methods in the literature for interpolating Lagrangian particles in continuous phase space \citep[e.g.,][]{Bai10,Laibe12a,Laibe12b,Pablo15,Yang16}. Different methods affect both the spatial accuracy and spatial resolution of final outcome. Generally, Lagrangian particles with a larger effective size reduces the spatial resolution, but gives better accuracy. Investigating which kind of method works best with SSA is one future direction of this work.

Another approach is to simulate the dust and gas both as grid-based fluids, with the dust being a pressure-less fluid. This is also a common approach in the literature \citep[e.g.,][]{Paardekooper06,Fu14,Laibe14,Lin17}. In this case, both the dust and the gas can use the SSA algorithm and evolve synchronously. While \secref{sec:SSA} expressed SSA in a Lagrangian approach, it is in fact straightforward to translate it to an Eulerian grid. As mentioned previously, the backbone of SSA is the drift-kick-drift leapfrog algorithm, which means that the drift steps (Equations \ref{eq:SSA_drift1} and \ref{eq:SSA_drift2}) naturally translate to advection steps in an Eulerian grid. In other words, the SSA algorithm already separates advection and forcing by operator splitting, as is commonly done in numerical fluid dynamics.

In an advection step, the mass flux across a cell boundary integrated over one time step can be expressed as:
\begin{equation}
    F = \int_{x_{\rm b}-v_{\rm ad}\,\dt}^{x_{\rm b}} \rho(t,\,x)\, A \, {\rm d}x \, ,
    \label{eq:advect}
\end{equation}
where $x_{\rm b}$ is the position of the boundary, $v_{\rm ad}$ is the advection speed, $A$ is the area of the boundary, and $\rho(t,\,x)$ is a continuous density function that is reconstructed from the discretized densities in the neighbouring cells. In an Eulerian grid, we can replace \eqnref{eq:SSA_drift1} with \eqnref{eq:advect} by using half a time step, choosing $t=t_{\rm i}$, and evaluating $v_{\rm ad}$ as
\begin{equation}
    v_{\rm ad} = \frac{1}{v_{\rm b}\,\dt/2}\int_{x_{\rm b}-v_{\rm b}\,\dt/2}^{x_{\rm b}} v(t_{\rm i},\,x) \, {\rm d}x \, ,
    \label{eq:v_ad}
\end{equation}
where $v$, like $\rho$, is the spatially reconstructed velocity, and $v_{\rm b}$ is the velocity at the cell boundary. \eqnref{eq:SSA_drift2} can also be transformed similarly, but using the updated $v$ and $\rho$ at $t=t_{\rm i+1}$. Momentum and energy fluxes are evaluated in the same way, with $\rho$ in \eqnref{eq:advect} replaced by the momentum and energy density, respectively.

An essential ingredient of this approach is the reconstruction method, which sets the the spatial accuracy and is often the determining factor in how numerically diffusive the method is. In this sense, SSA is only half the complete algorithm. For the forcing steps, Equations \ref{eq:SSA_staggered} and \ref{eq:SSA_kick} can be applied directly by using the cell-center values of $v$, $f$, $\ts$, and $\vg$. One problem, however, is that $v_{\rm g,1}$, the gas velocity at $t=t_{\rm i}+\dt/2$, is needed at the staggered step, but that information is not available since the gas also requires the dust velocity at $t=t_{\rm i}+\dt/2$ in order to evolve. Some way to approximate the midpoint velocities is therefore necessary.

To account for back-reaction on the gas, one could simply reverse the roles of $v$ and $\vg$, or strictly enforce momentum conservation by adding to the gas the momentum lost by the dust due to drag. The latter choice may have better conservative properties, but may not be more accurate in general. In principle, one could even extend SSA to multiple species by setting up an eigenvalue problem for momentum exchange \citep[e.g.,][]{Pablo19} and deriving semi-analytic expressions analogous to \ref{eq:SSA_kick}. Decisions surrounding how to perform spatial reconstruction in the advection steps, how to approximate the midpoint velocities in the forcing steps, and how to include back-reaction, all need to be carefully considered and tested in order to design a grid-based hybrid code using SSA.

\acknowledgements
{\small \noindent We thank Eugene Chiang for useful suggestions and discussions. We also thank an anonymous referee for helpful and encouraging comments. This work was performed under contract with the Jet Propulsion Laboratory (JPL) funded by NASA through the Sagan Fellowship Program executed by the NASA Exoplanet Science Institute.}

\appendix
Following the example of equation \eqnref{eq:v_ana}, we can write an analytic solution for the evolution of velocity for a quadratic drag law, such as the supersonic drag described by \eqnref{eq:super}, subject to an external force parallel to the direction of the drag. If we assume that the stopping time $t_s$, velocity scale $c$, gas velocity $v_g$, and external force $f$ are constant throughout a time step, and taking $k \equiv {\rm sign} (v - v_g)/c t_s$, we obtain the following five expressions in different cases, analogous to those derived in \cite{Han16}. We note that our equation \eqref{eq:stokes_no_force} is equivalent to \eqref{eq:super_ana}, but rearranged for ease of comparison with the other expressions:
\begin{subequations}\label{eq:app_quadratic}
\begin{equation}
    v_{-+}(t, f, k, v_0) = v_g + \frac{(v_0 - v_g) + \sqrt{|f/k|} \tanh{(\sqrt{|fk|}t}) }{1 + \sqrt{|k/f|} |v_0 - v_g| \tanh{(\sqrt{|fk|}t})} \, , \ \  (k < 0, f > 0) \, ;
\end{equation}
\begin{equation}
    v_{+-}(t, f, k, v_0) = v_g + \frac{(v_0 - v_g) - \sqrt{|f/k|} \tanh{(\sqrt{|fk|}t}) }{1 + \sqrt{|k/f|} |v_0 - v_g| \tanh{(\sqrt{|fk|}t})} \, , \ \  (k > 0, f < 0) \, ;
\end{equation}
\begin{equation}
    v_{--}(t, f, k, v_0) = v_g + \frac{(v_0 - v_g) - \sqrt{|f/k|} \tan{(\sqrt{|fk|}t}) }{1 + \sqrt{|k/f|} |v_0 - v_g| \tan{(\sqrt{|fk|}t})} \, , \ \  (k < 0, f < 0) \, ;
\end{equation}
\begin{equation}
    v_{++}(t, f, k, v_0) = v_g + \frac{(v_0 - v_g) + \sqrt{|f/k|} \tan{(\sqrt{|fk|}t}) }{1 + \sqrt{|k/f|} |v_0 - v_g| \tan{(\sqrt{|fk|}t})} \, , \ \  (k > 0, f > 0) \, ;
\end{equation}
\begin{equation}\label{eq:stokes_no_force}
    v_{\pm 0}(t, f, k, v_0) = v_g + \frac{v_0 - v_g}{|v_g - v_0| |k|t + 1} \, ,  \ \  (f = 0) \, .
\end{equation}
\end{subequations}

Complication arises when the sign of $k$, or equivalently $(v - v_g)$, changes within a time step, which can happen when $v_{++}$ or $v_{--}$ is the appropriate solution at the start of the time step. This discontinuity is not necessarily physical. In reality, the quadratic drag would likely become linear as the relative speed $(v - v_g)$ approaches zero. Nonetheless, for quadratic drag only, one way to account for this sign change is to implement the following switch:
\begin{subequations}
\begin{equation}
    v_{\rm i+1} = 
\begin{cases}
    v_{++}(\Delta t, f_{\rm i}, k_{\rm i}, v_{\rm i}) & \, , k_{\rm i}, f_{\rm i} > 0, \Delta t < \Delta t_{++} \, ; \\
    v_{-+}(\Delta t - \Delta t_{++}, f, k_{\rm i}, v_{\rm i}) & \, , k_{\rm i}, f_{\rm i} > 0, \Delta t \geq \Delta t_{++} \, ;
\end{cases}
\end{equation}
\begin{equation}
    v_{\rm i+1} = 
\begin{cases}
    v_{--}(\Delta t, f_{\rm i}, k_{\rm i}, v_{\rm i}) & \, , k_{\rm i}, f_{\rm i} < 0, \Delta t < \Delta t_{++} \, ; \\
    v_{+-}(\Delta t - \Delta t_{--}, f, k_{\rm i}, v_{\rm i}) & \, , k_{\rm i}, f_{\rm i} < 0, \Delta t \geq \Delta t_{--} \, ;
\end{cases}
\end{equation}
\begin{equation}
    v_{\rm i+1} = 
\begin{cases}
    v_{-+}(\Delta t, f_{\rm i}, k_{\rm i}, v_{\rm i}) & \, , k_{\rm i} < 0, f_{\rm i} > 0 \, ; \hphantom{\Delta t \Delta t_{--}} \\
    v_{+-}(\Delta t, f_{\rm i}, k_{\rm i}, v_{\rm i}) & \, , k_{\rm i} > 0, f_{\rm i} < 0 \, ; \hphantom{\Delta t \Delta t_{--}} \\
    v_{\pm 0}(\Delta t, f, k_{\rm i}, v_{\rm i}) & \, , f_{\rm i} = 0 \, ;
\end{cases}
\end{equation}
\end{subequations}
where $\Delta t_{++}$ and $\Delta t_{--}$ are the times at which $v_{++} - v_g$ and $v_{--} - v_g$ change in sign:
\begin{subequations}
\begin{equation}
    \Delta t_{\rm ++} = \sqrt{|fk|}^{-1}\arctan\left[-\frac{(v_0 - v_g)}{\sqrt{|f/k|}}\right] \, ,
\end{equation}
\begin{equation}
    \Delta t_{\rm --} = \sqrt{|fk|}^{-1}\arctan\left[\frac{(v_0 - v_g)}{\sqrt{|f/k|}}\right] \, .
\end{equation}
\end{subequations}
Using the velocity update above, and the simple position update $x_{\rm i+1} = v_{\rm i+1} \Delta t$, we have by construction a semi-analytic method for quadratic drag. The method is unconditionally stable; when $\tau_{\rm quad} \gg 1$, $ v_{\rm i+1} \rightarrow v_g \pm \sqrt{|f_{\rm i}/k_{\rm i}|}$, depending on whether the sign of $(v_{\rm i} - v_g)$ was positive or negative respectively and whether velocity underwent a sign-change during a time step.

As mentioned before, the scheme is exact when the external force $f$, gas velocity $v_g$, and drag coefficient $k$ are constants (up to sign change in $k$), and when $\vec{f}$ is parallel to $(\vec{v}_0 - \vec{v}_g)$. For general and possibly non-parallel forces (replacing $f$ with $f_{\parallel}$ or $f_{\bot}$, $(v_0 - v_g)$ with $(\vec{v}_0 - \vec{v}_g)_{\parallel}$ or $(\vec{v}_0 - \vec{v}_g)_{\bot} \equiv 0$, and $|v_0 - v_g|$ with $|\vec{v}_0 - \vec{v}_g|$ in \eqref{eq:app_quadratic}), the convergence of the scheme degrades to $1^{\rm st}$ order, and reduces to symplectic Euler for $\tau_{\rm quad} \ll 1$. We note that unlike with linear drag, the parallel velocity influences the perpendicular quadratic drag force and vice versa, making a true multi-dimensional analytic solution intractable \citep{Han16}.

\bibliographystyle{yahapj}
\bibliography{Lit}

\end{document}